# Picoscale Magnetoelasticity Governs Heterogeneous Magnetic Domains in a Noncentrosymmetric Ferromagnetic Weyl Semimetal

Bochao Xu[1,2], Jacob Franklin[1,2], Hung-Yu Yang[2], Fazel Tafti[2] and Ilya Sochnikov[3,*]

[1]Physics Department, University of Connecticut, Storrs, CT USA, 06269

[2]Department of Physics, Boston College, Chestnut Hill, MA 02467, USA

[3]Institue of Material Science, University of Connecticut, Storrs, CT USA, 06269

[*]Corresponding author: ilya.sochnikov@uconn.edu

**Magnetic Weyl semimetals are predicted to host emergent electromagnetic fields at heterogeneous strained phases or at the magnetic domain walls. Tunability and control of the topological and magnetic properties is crucial for revealing these phenomena, which are not well understood or fully realized yet. Here, we use a scanning SQUID microscope to image spontaneous magnetization and magnetic susceptibility of CeAlSi, a noncentrosymmetric ferromagnetic Weyl semimetal candidate. We observe large metastable domains alongside stable ferromagnetic domains. The metastable domains most likely embody a type of frustrated or glassy magnetic phase, with excitations that may be of an emergent and exotic nature. We find evidence that the heterogeneity of the two types of domains arises from magnetoelastic or magnetostriction effects. We show how these domains form, how they interact, and how they can be manipulated or stabilized with estimated lattice strains on picometer levels. CeAlSi is a frontier material for straintronics in correlated topological systems.**

## Introduction

Topological materials, namely the materials whose electronic structures are protected by non-trivial symmetries, have drawn great interest in the past decade [1,2]. The early discovered topological insulators, with robust conducting surface states induced by the topology of the bulk band structure, have led to the realization of various novel phenomena such as the quantum anomalous Hall effect.[3–6] With the more recent discovery of Dirac Weyl fermions in semimetals, research focus has shifted to topological semimetals.

Topological Weyl semimetals (WSM) are characterized by the linear dispersion around the band crossing points, termed the Weyl nodes.[7–11] Their non-closed surface states, the Fermi arcs, and the band crossing have been experimentally observed through angle-resolved photoemission spectroscopy (ARPES).[12,13] Weyl nodes are protected by breaking either spatial-inversion symmetry or time reversal symmetry. Magnetic Weyl semimetals, such as antiferromagnetic semimetal $Mn_3Z$ [14,15] and kagome ferromagnetic semimetal $Co_3Sn_2S_2$ [16,17], are particularly interesting since their topological states are easily tunable using an external magnetic field. Experimental studies show that magnetic Weyl semimetals with broken time reversal symmetry may generate strong anomalous Hall effects due to their large Berry curvature.[16,18] The Nernst effect was also shown to be connected to the Berry curvature.[19,20] Moreover, the magnetic textures, such as magnetic domain walls, in such materials can induce localized charges and be effectively treated as axial gauge fields.[21–23] Magnetic domain walls, as a result of topological defects in the real-space, may largely affect the electronic behaviors in momentum-space.[24] A closer look at domain walls at mesoscopic level is desirable to understand their dynamics, and it may lead to a practical pathway to unveil the interplay between Weyl electrons and magnetic textures.

Recently, the RAlX (R=Rare earth elements, X=Ge or Si) family of compounds emerged in studies of magnetism in Weyl semimetals as they were proposed to be potential candidates of magnetic Weyl semimetals with remarkable tunability.[25,26] Depending on the choice of rare earth ions, RAlX can be ferromagnetic, antimagnetic or nonmagnetic and hosts either type I/or type II Weyl fermions. PrAlGe was expected to be a type I magnetic WSM and had indirect evidence of hosting nanoscale ferromagnetic domain walls which are tunable by changing the temperature and external magnetic field.[23] In addition, CeAlGe was experimentally proven to be a type II antiferromagnetic WSM with a phase transition at 5K.[27,28] A singular angular magnetoresistance study performed on CeAlGe also shows the existence of controllable high-resistance domain walls, which result from broken magnetic symmetry and fermi surface mismatch in a nearly nodal electronic structure.[29] CeAlSi, which has the same $I4_1md$ space group, was recently shown to be a noncentrosymmetric ferromagnetic WSM and exhibit an anisotropic anomalous Hall effect with different behaviors when measured along the easy- and hard- axes.[30] The large ferromagnetic domains observed in CeAlSi below $T_c$=8.3K in ref.[30] suggest that the interplay between magnetic order and topological band structure may be responsible for a loop-shape Hall effect observed with a magnetic field along the hard-axis. In this paper, we further study of the different magnetic domains in this material by investigating their origin and showing ways to manipulate them.

In this paper, we address a question of whether we understand the magnetic behavior of the ferromagnetic Weyl systems. Here, we study magnetic domains in CeAlSi crystals by investigating the origin of their heterogeneity and showing ways to manipulate them. The results below will guide future efforts in building quantum devices from these materials in order to observe topological effects. Surely, considerations of the physics of the domains observed here will be important in this endeavor.

## Experimental

We report imaging of magnetic domains in CeAlSi using scanning Superconducting QUantum Interference Device (SQUID) microscopy (SSM). SSM is one of the most sensitive tools to measure tiny magnetic features near a sample surface. In our sensors, the gradiometric sensing pick-up loops are integrated with the well-shielded body of the SQUID and are concentric with the micro-coils, which allows us to apply a local magnetic field to the sample [31–33]. Thus, our sensors can measure and image magnetic susceptibility in addition to spontaneous magnetization or dc magnetic flux. The ability to image magnetic ac susceptibility allows us to discriminate between domains with different magnetic characters as shown below, which otherwise look practically indistinguishable in the magnetic dc signal.

In this experiment, we use a SQUID [31] which has a niobium pick-up loop with an inner diameter of 3 μm and an outer diameter of 3.5 μm and a one-turn field coil with an inner diameter of 6 μm and an outer diameter of 7 μm. Spatial resolution of this sensor is about 3-6 μm depending on the sample features (sharp changes can be resolved with better resolution). The SQUID was attached to a custom-built piezoelectric scanner. The SQUID, scanner, and samples were cooled inside a Montana Instruments Fusion closed-cycle cryostat similar to the one described by Shperber *et al* [34].

CeAlSi crystals were grown using a self-flux method [35]. The starting materials Ce, Al, and Si were mixed in a mole ratio of 1:10:1 and placed in an alumina crucible inside an evacuated silica tube. The mixture was heated to 1000 °C for 12 h, cooled to 700 °C at 0.1 °C/min, annealed at 700 °C for 12 hours and then centrifuged to decant the residual Al-flux. The crystals are several milliliters wide with metallic luster, and they are stable in air. The CeAlSi samples have a ferromagnetic transition temperature of ~8.3 K. We polished the [001] surface of the samples using a standard polishing jig and a diamond lapping film with a 0.5 micron grade to obtain a flat surface suitable for the scanning SQUID imaging. $I4_1md$ space group structure and the ferromagnetic phase transition were previously verified by optical second harmonic generation and neutron scattering [35]. In this work, magnetostriction was measured by applying a small field in a <1 1 0> direction to a sample glued to a strain gauge.

## Results

Figure 1 A shows typical domains in the magnetization; also visible are in-phase and out-of-phase ac magnetic susceptibility components. The DC flux images display the stray field piercing the sample surface parallel to the *a - b* crystallographic plane. The in-phase and out-of-phase susceptibility show *two* types of domains with different strengths of response. The in-phase susceptibility shows that these domains differ in the strength of paramagnetic-like response to local field of about 0.3 G at a typical frequency of 1800 Hz. The increased out-of-phase susceptibility in some domains correspond to those domains exhibiting energy losses (hysteresis) on the timescales of the local excitation ac magnetic field (note, the negative sign of the out-of-phase component is a matter of convention[36,37]).

We hypothesized that the different domains arise due to magnetoelastic and magnetostriction effects that modify some of the regions of the sample. These regions experience larger strains due to the internal magnetization interacting with the lattice; see Figure 1 B. To confirm that there is such a lattice-field connection in CeAlSi crystals (see sample A shown in Figure 1 C), we attached one of the crystals to a miniature 1 mm long strain gauge positioned inside a small solenoid along with three other gauges comprising a Wheatstone bridge. This resulted in the high accuracy resistance measurements required to detect the tiny magnetostriction effects. We observed a magnetoelastic response from the sample at low temperatures below the ferromagnetic transition (see Figure 1 D). The magnetostriction response is small but clearly detectable for the limited fields we applied: on the order of $10^{-5}$ at a few tens of Gauss in-plane fields. This strain corresponds to single nanometers of the total sample length changes.

As we show in Figure 2, such small external fields suffice to substantially change the landscape of the domains. In this figure, field cooled data is shown for representative field values ranging from a single Gauss to a couple of tens of Gauss. At lower fields, a substantial fraction of metastable domains is apparent alongside small stable domains. Stable FM domains were enhanced with presence of larger external fields. The average area of *stable* domains grows to more than twice the zero-field domain areas, as we apply a field of just over 10 G.

Applying external field is not the only method that can be used to manipulate the domains. By cycling the sample through the transition several times in zero field, but every time lowering the temperature to which the sample was warmed up until it reached 6 K, we were able to train the domains even without the external field. Figure 3 shows the magnetic landscapes of both the zero-field-cooled sample and the sample treated by the training process. The ferromagnetic phase was modified due to the considerable growth of the stable FM domains after the sample was trained. Figure 3 shows the histogram of the stable domains. This annealing confirms that the domains with smaller susceptibility are the thermodynamically more stable ones; hence the other domains were defined as 'metastable'.

We illustrated the FM phase transition itself in CeAlSi by showing the temperature evolution of FM domains in Figure 4. The presented sample A (also see sample B in Figure A 6) was first zero field cooled down to 6 K and then warmed to over 9 K with images taken at several temperatures between. Metastable domains, identified by light marine (higher signal) and light green (higher negative signal) areas in the in-phase susceptometry and the out-of-phase susceptometry, grew as the temperature increased. Susceptibility signals reached a maximum value at temperatures near $T_c$, where both stable and metastable domains were no longer distinguishable, and then eventually turned into narrow features along the lower temperature domain boundaries. No strong magnetic signals or FM domains were observed at 8.26 K, consistent with the bulk transition temperature around 8.3 K [35]. In addition, the observed domain boundaries are within 3 μm sharpness, which is supportive of vertical walls.

Further small-area were repeatedly scanned without an external field, and showed an unusual slowly-fluctuating behavior in the out-of-phase susceptometry at 6 K, see Figure 5. The wave-like features across the metastable domain were found to change their shape over the timescale of minutes and the slow-dynamics persisted over hours. The persistent dynamics of the out-of-phase component is surprising given that all the susceptibility measurements are done in the linear response regime (no dependence on the ac field amplitude, see Appendix). Typically, what fluctuates in magnetic systems is magnetization, but not susceptibility [38,39]. Furthermore, we find that the so-called Cole-Cole plots (see Figure A 3) of the in-phase *vs.* out-of-phase components of the magnetic susceptibility show short, incomplete arcs that are a signature of a wide distribution of activation times in a frustrated system [36,37,40].

## Discussion

The DC magnetic flux images are consistent with the results from bulk measurements [35] in terms of the direction of magnetization being in-plane (normal to the *c*-axis). The DC magnetic flux images show that the stray field penetrates out of the *a*-*b* surface at the in-plane domain boundaries. We found that zero-field-fast-cooled domains are rather small (a few microns in-size) and a weak in-plane field of a few Gauss is enough to develop very large in-plane FM domains (a few hundreds of microns in size). The typical level of the observed DC flux variations is a few $\Phi_0$, which is consistent with the remnant magnetization determined from the *c*-axis bulk magnetization measurements [35]. The remnant *a*-axis magnetization predicts DC signal on the order of hundreds of $\Phi_0$ if the domains were to be polarized out-of-plane (note that we could not invert the DC flux images reliably to determine the exact in-plane configuration due to the intricacy of the patterns).

Thus, the magnitude of the signal is consistent with the in-plane magnetization in the stable domains.

In our experiments, the in-phase *c*-axis AC susceptibility channel probes small canting of the in-plane atomic magnetic moments in the stable FM domains, which exhibit very small *c*-axis signal at 6 K. Domains identified as metastable showed larger responses, likely due to a magnetically disordered frustrated state, or due to under-resolved finer structure to be studied in future works (such as skyrmions [41], see the ripples in Figure 5). These domains' types could be distinguished in the susceptibility channel but not in the DC magnetization, highlighting the power of our susceptibility imaging technique. The location of the domains is random with temperature cycling through the transition which rules out chemical phase separation. Interestingly, domain walls exhibited signal in the out-of-phase component, indicating that the metastable domains and domain walls have some intimate connection of yet undetermined origin.

The metastability is likely associated with the interlayer anti-ferromagnetic-like (partial) ordering. Our neutron scattering and bulk magnetization data [35] show that the magnetization lies in the *a-b* crystallographic plane. In the proper ferromagnetic phase, in-plane Ce magnetic moments within the unit cell are stacked in [$cos(45°+θ/2)$ $sin(45°+θ/2)$ 0] and [$cos(45°-θ/2)$ $sin(45°-θ/2)$ 0] directions with the angle between the Ce moments estimated to be $θ=68°$ [35]. This stacking, when coarse-grained, produces the net in-plane magnetization in the [1 1 0] direction in this example (or other three degenerate directions in different proper domains). Under a built-up magnetoelastic strain, the *c*-axis length changes, and the interlayer interactions become modified. Hence, differently stacked configurations in terms of the Ce moment alternation may emerge, which we believe is the likely origin of the metastable domains that are identified in the susceptibility channels. Individually magnetized in-plane Ce layers can be organized in an antiferromagnetic-like sequence in the c-axis direction, or can be organized in disordered stacking, or even in a chiral (rotating) stacking order. Chiral magnetization can possibly even lead to skyrmions. This interpretation is fully consistent with our previous bulk results. These stacking orders would be impossible to distinguish in neutron scattering because even a small external magnetic field, as our current scanning SQUID work shows, can wipe out the regions with metastable stacking orders. Further, such non-ferromagnetic stackings might result in a fine in-plane structure, hence the suggestion about skyrmions.

The out-of-phase AC susceptibility measurements probe dissipation effects in the "movement" of the in-plane spins. They do not only probe mere canting. The temporally changing out-of-phase component must be accompanied by some hysteresis, and it is not enough to have simple canting of the spins. We notice that at low temperature, the stable FM domains show no out-of-phase component as one expects for simple, fully reversible canting. However, the metastable regions are now more clearly identified as those that exhibit non-zero out-of-phase signal. The out-of-phase component grows in magnitude with temperature. Thus, it appears to be intimately related to the mechanism of the phase transition (see also temperature dependence of the permeability in Appendix).

The magnetic phases of RAlX (R = rare-earth and X = Si, Ge) materials are known to be very sensitive to small *lattice variations* between the family members, even though the microscopic mechanism for such sensitivity is not known [42–47]. Our work shows evidence that such sensitivity to strain can occur within a specific material. We probed how the size of the samples change under small in-plane field and found that in terms of the lattice deformations such magnetostriction strains could correspond to *picometer level* displacements for the fields used here.

Note, the off-stoichiometry of Al and Si in our material is less than 2%, without any major changes in the magnetic behavior of the material as pointed out in ref. [35]. Bulk magnetization curves in that reference are identical between several samples grown under different conditions, but the Hall effect shows mild variations because a slight off-stoichiometry can shift the Fermi-level. However, note that magnetism comes from the *f*-electrons of Ce that are 6 eV below the Fermi-level. Thus, a change of Fermi level by a few meV has no impact on the localized moments responsible for magnetism in CeAlSi.

The internal coarse-grained magnetization corresponds to a higher internal magnetic field: larger than our test fields in Figure 1 D. We estimate the internal field (magnetic flux density, **B**) to be on the order of 1000 G (from the bulk results [35]), which means that the built-in strain in or around the domains is likely an order of magnitude or two larger than the strain measured here in the ~30 G external field. Estimated $10^{-3}$ strains at internal fields of ~1000 G would correspond, then, to an order of a picometer distortion per unit cell.

Upon formation of internal magnetization in the stable ferromagnetic domains, the domains could build up internal strain due to the magnetostriction or magnetoelastic effects [48,49] and effectively apply a somewhat larger strain at the surrounding boundaries, producing a different magnetic phase around them due to the excessive strain (surface-origin of the signal is ruled out based on the approach curves, see Figure A 4). We believe that this is the reason for the observed magnetic phase separation represented by domains with small and large magnetic AC susceptibility: one has large uniform domains, and another is either frustrated or a fine-structure matrix modified by strain.

The out-of-phase component of the susceptibility is related to magnetization noise via the fluctuation dissipation theorem. In our case the out-of-phase component is itself "noisy" in the metastable domains, which is highly unusual and is probably a reflection of some magnetic frustration in the magnetic state. To the best of our knowledge the only case where susceptibility was found to be fluctuating is a system exhibiting superconductor to insulator transition, with slow quantum fluctuations in the diamagnetism [50]. We speculate that the two material systems might be related through the role of frustration: frustration in the lattice of spins in the magnetic system and frustration in the superconducting vortices in the superconducting system.

What this combination of stable and metastable phases offers are the opportunities for combining different topologically nontrivial interfaces in different magnetic domains of the noncentrosymmetric Weyl semimetal. The metastable domains exhibit magnetic fluctuations that may be of magnetically non-trivial nature by themselves, as expected in other materials [51,52]. Further, the coexistence of these phases may be related to the puzzling Hall effect signals in these materials [35]. We envision the facilitation of emergent electromagnetics [53] by controlling these phases and the interfaces between them with magnetic field and strain knobs. While we do not claim to have demonstrated the ultimate electronic topological effects in this work, our result is a critical milestone towards this goal. Simply put, we need to know how to control and manipulate magnetic domains and boundaries in devices made from this material. The current work provides a keystone playbook for these future experiments in CeAlSi and other related materials.

Future studies should focus on measuring the expression of these magnetic states in the electronic properties, and on fine-tuning these phases with external strain. Building devices that will utilize the domains and domain walls to uncover emergent electromagnetic behavior becomes a more realistic target with our findings: We show how magnetic domains form in CeAlSi, how they interact, and how they can be manipulated or stabilized with lattice strains estimated to be on picometer levels, which means it should be possible to induce such small strains externally using a strain cell and perhaps even enhance the ferromagnetic transition temperature [54–57]. Furthermore, the discovered metastable domains have a potential of hosting skyrmions due to the noncentrosymmetricity. Domain physics in Weyl semimetals is important for realizing many new functionalities in practical devices with tunability potentially useful for applications ranging from yet to be developed topological straintronics [58], to terahertz devices [59,60], to even qubits [61].

## Summary

We have performed a comprehensive scanning SQUID study of single crystals of the ferromagnetic noncentrosymmetric Weyl semimetal candidate CeAlSi. We discovered a heterogenous state that is formed by the proper ferromagnetic in-plane domains and a type of metastable domains characterized by energy dissipation. The latter forms most likely due to the magnetostriction or magnetoelastic effects and enhanced strain around the ferromagnetic domains, surrounded by the matrix of a possibly magnetically frustrated configuration. Coexistence of such spatially separated strained phases will allow for the engineering of topologically nontrivial domain boundaries with possible emergent electrodynamics and new functionalities in future works.

## Acknowledgments

We thank J. R. Kirtley for the inductance calculation software code, the State of Connecticut for financing construction of the scanning SQUID, the College for Liberal Arts and Sciences of the University of Connecticut for special graduate research assistantship support for B. X. and J. F. I. S. acknowledges the US DOD for partial support. The work at Boston College was funded by the National Science Foundation under grant no. DMR-1708929.

## Data availability

The data sets presented in the present study are available from the corresponding author upon reasonable request.

## Author Contributions

B. X., J. F. and I. S. performed scanning SQUID imaging experiments, H.-Y. Y. and F. T. synthesized the crystals and performed their characterization. All authors contributed to the manuscript preparation.

## Competing Interests

The authors declare no competing financial or non-financial interests

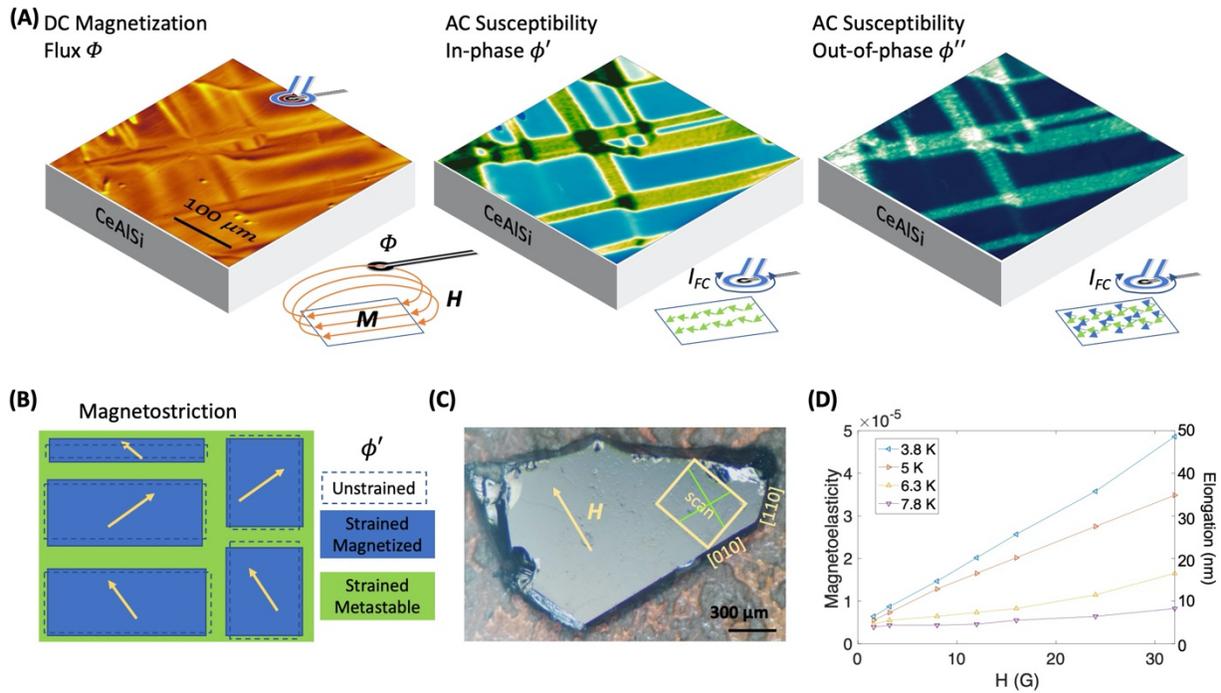

*Figure 1. Main experimental details of imaging heterogeneous magnetic domains in CeAlSi and their relation to magnetostriction effects. (A) Scanning SQUID sensor shown schematically as a small pickup loop and field coil that measures DC magnetic flux and AC magnetic susceptibility. Shown are images at 6K, which is below the ferromagnetic phase transition of 8 K. Large in-plane domains are visible in the DC flux images. AC in-phase susceptibly shows two types of domains characterized by low (blue) and high (green) susceptibility. The out-of-phase susceptibility image shows no energy losses (dark green) in the low susceptibility regions and appreciable energy losses (light green) in the high susceptibility regions. These are stable and metastable domains, correspondingly. (B) A schematic explaining how the metastable regions form in the material as local strains are induced by the magnetostriction in the in-plane ferromagnetic domains. (C) One of the single crystal samples imaged in this work. The orientation of the external field pertaining to the data shown below is shown here. Green lines are typical orientation of the domains in the scan area shown by the beige rectangular frame. (D) Strain induced by weak in-plane field as measured using a strain-gauge method. The magnetoelasticity drops with an increase in temperature, consistent with the ferromagnetic phase transition temperature of ~8.3 K. The absolute sample elongation (per 1 mm length of the gauge) is shown to be very small (in the nanometers range,*

*right ordinate). The strain values are very small, corresponding to what we estimate is less than a picometer deformation of the unit cell, yet they can play important role as shown in this work.*

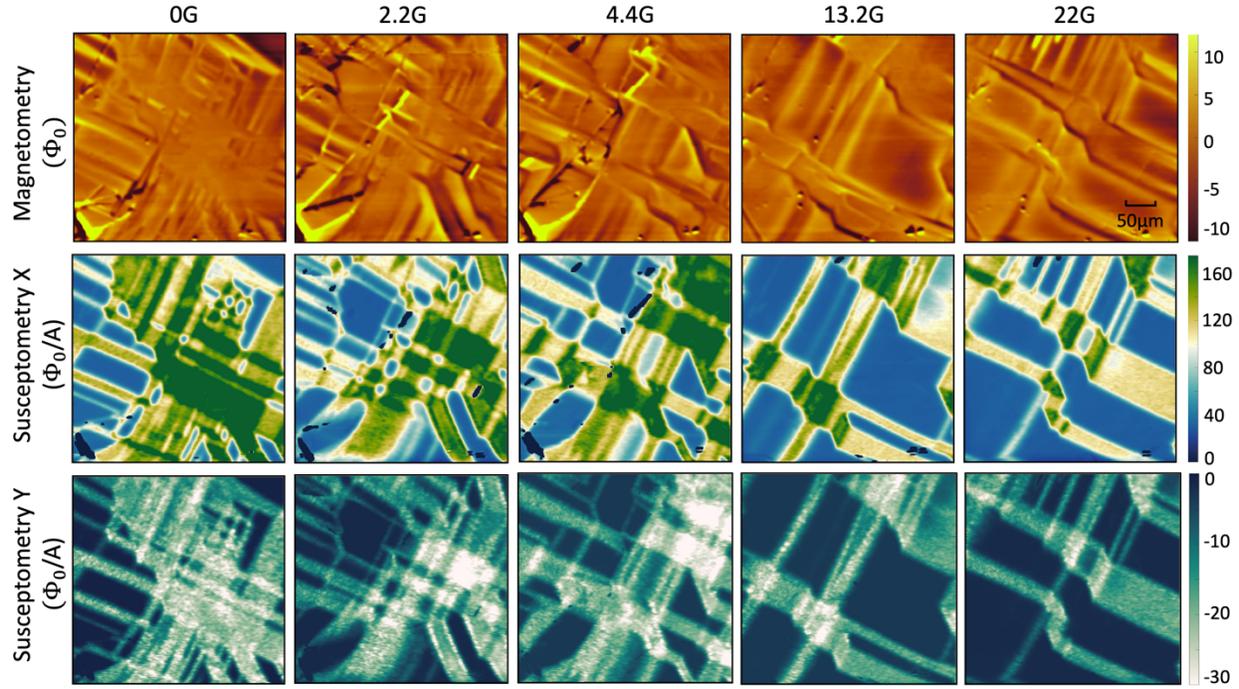

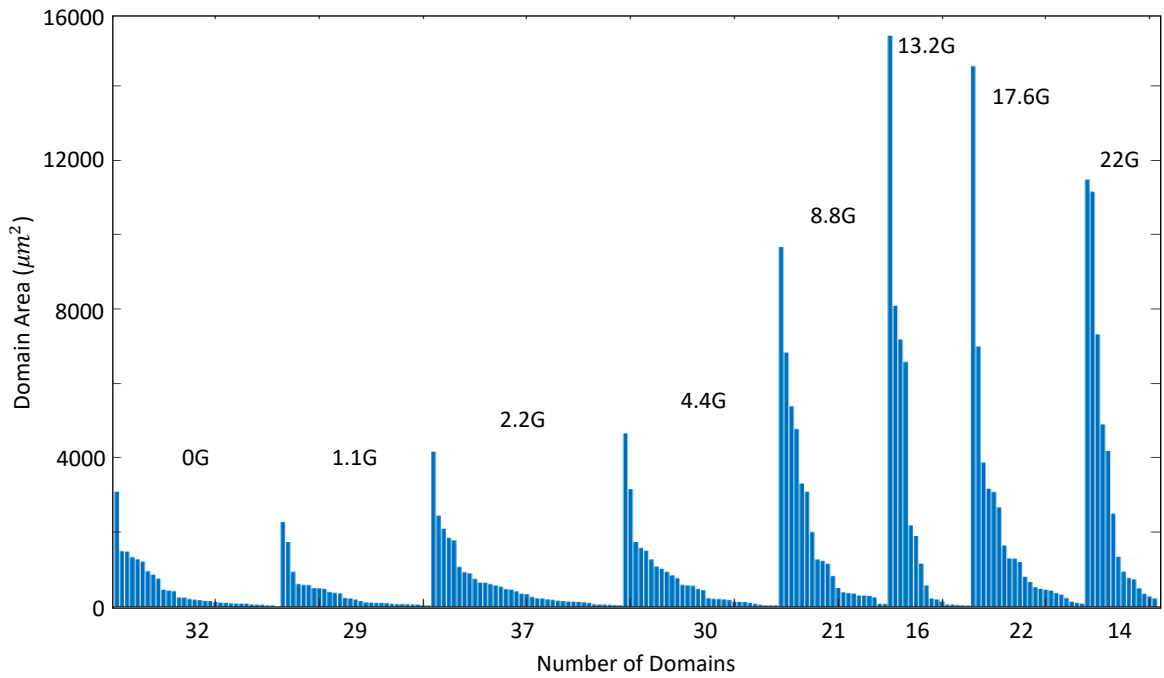

*Figure 2. The growth of domains with magnetic field in a CeAlSi crystal that was as-cooled through the transition at ~8K. (top panel) Upper row: dc magnetizations in flux units. Middle and lower rows: the in-phase component and the out-of-phase component of the ac susceptibility. Dark green and dark marine areas in the in-phase and out-of-phase components, respectively, correspond to stable in-plane*

*domains. Light green and light marine areas in the in-phase and out-of-phase components, respectively, are metastable domains. The domains are lined-up with the crystallographic <1 1 0> directions. (bottom) The histograms quantify how the stable domains grow with field, while the metastable domains shrink (see also Figure A 5).*

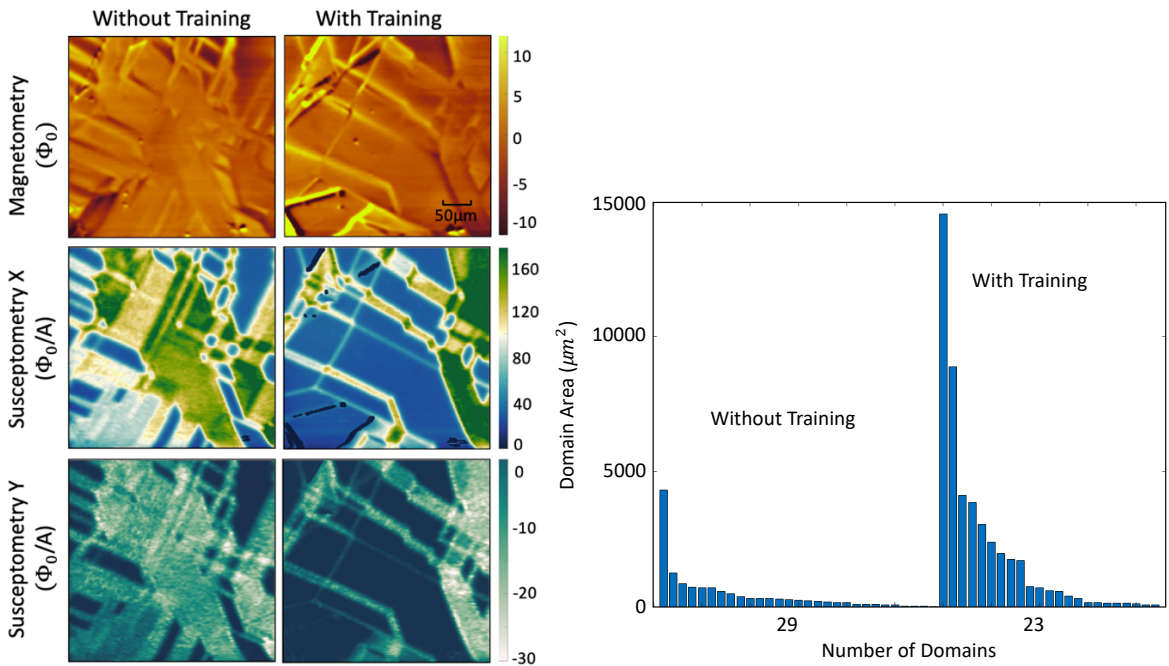

*Figure 3. Growth of magnetic domains by means of training. First column: as-cooled zero-field images. Second column: sample cooled in a step-like fashion through the transition. The lower susceptibility domains can be grown with this annealing method. The histogram quantifies the growth of the stable domain. This annealing experiment identifies the domains with lower susceptibility as stable and with higher susceptibility as metastable.*

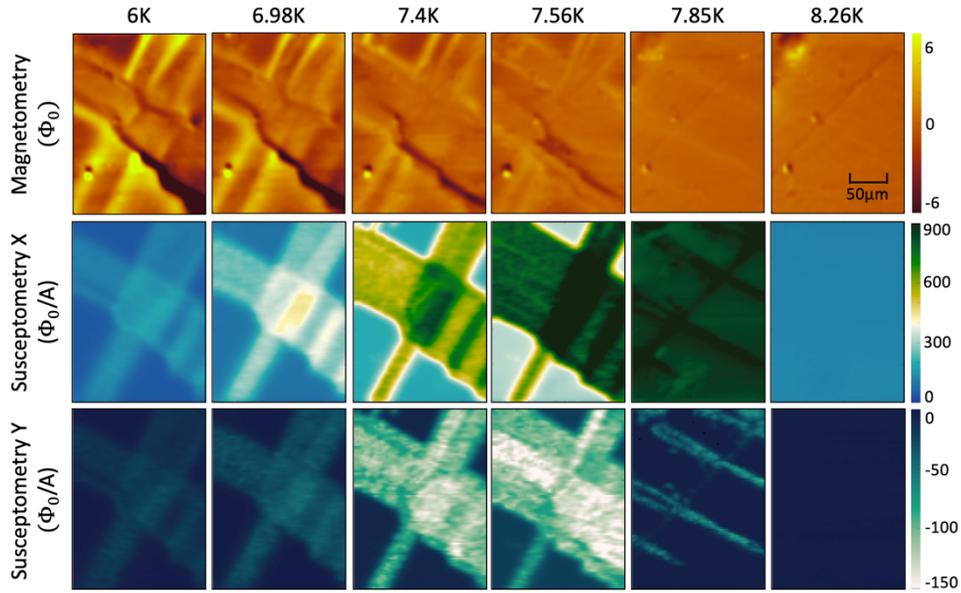

*Figure 4. The temperature evolution of the magnetic domains in the CeAlSi crystal. Upper row: dc magnetizations in flux units. Middle and lower rows are the in-phase component and the out-of-phase component of the ac susceptibility. Dark green and dark marine areas in the in-phase and out-of-phase components, respectively, are stable in-plane domains. Light green and light marine areas in the in-phase and out-of-phase components, respectively, are metastable domains. The stable domains shrink, while the metastable grow towards the transition, and eventually collapse to narrow features reminiscent of the domain boundaries roughly lined up with the crystallographic <1 1 0> direction. The dc magnetization contrast reduces towards the transition, while the amplitude of the in-phase and out-of-phase signals grow. The metastable regions show enhanced signals. The ~8.3 K transition temperature is in agreement with the bulk results [35]. Above the bulk transition temperature of 8.3 K, the bulk ferromagnetic domains disappear. Some small paramagnetic (not ferromagnetic) signal persists for another ~0.5-1 K, see Figure A 1. These might be fluctuations above the critical temperature.*

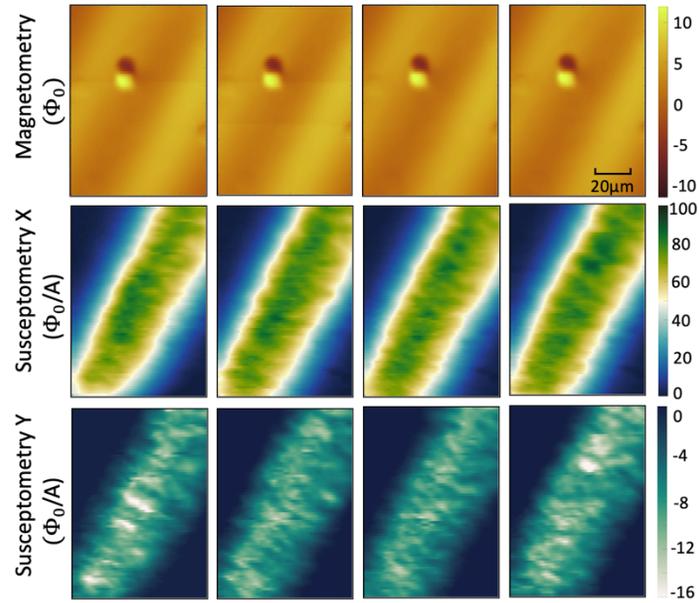

*Figure 5. Slow fluctuations in the out-of-phase susceptibility across one of the metastable domains. Time between images is about 10 min, each image has 75 lines, line-time 3.1 sec. If one follows the whiter spots, they can be noticed to change from one image to another, while the apparent size of the features of the wavy pattern is similar. Such fluctuations are extremely unusual and indicate a possible switching mechanism in a magnetically frustrated system.*

# Appendix

## Sample A

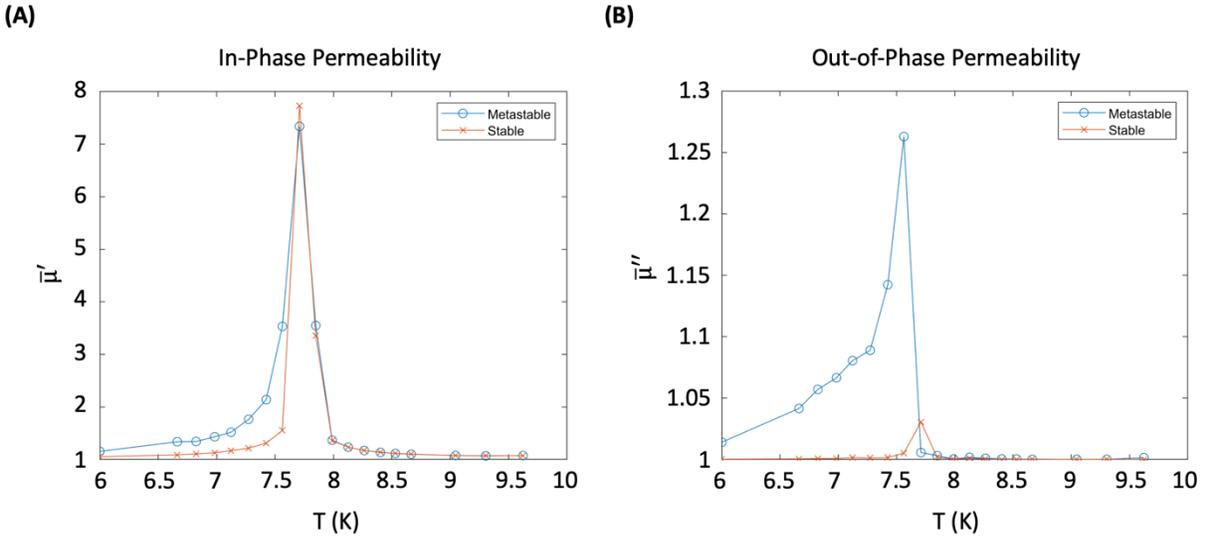

*Figure A 1. Permeability of the sample in a stable and a metastable domain. The permeability was extracted by fitting the Kirtley-Kogan model to the approach curves [62]. Touchdowns were performed in two locations: one in a metastable domain, and one in a stable domain. The normalized permeability, $\bar{\mu} = \mu/\mu_o$, is one of the fit parameters and is extracted from the fits. (A) The in-phase normalized permeabilities as a function of temperature for both locations. (B) The out-of-phase normalized permeabilities as a function of temperature for both locations. Note, the sharp drop in $\mu''$ is likely due to the metastable domain turning stable before the bulk phase transition was reached.*

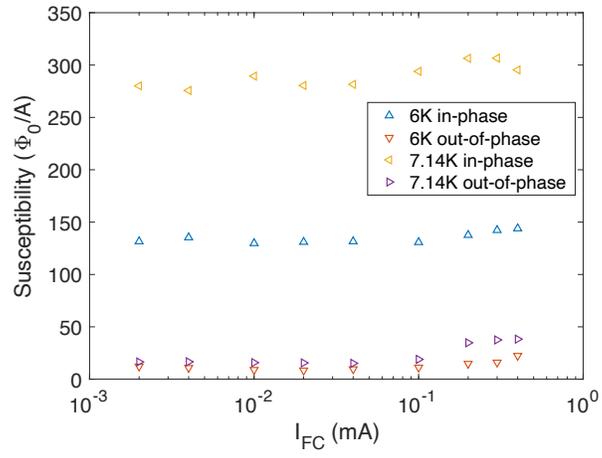

*Figure A 2. Susceptibility of a metastable region as a function of the field coil excitation current. The response is linear for amplitudes used in this work.*

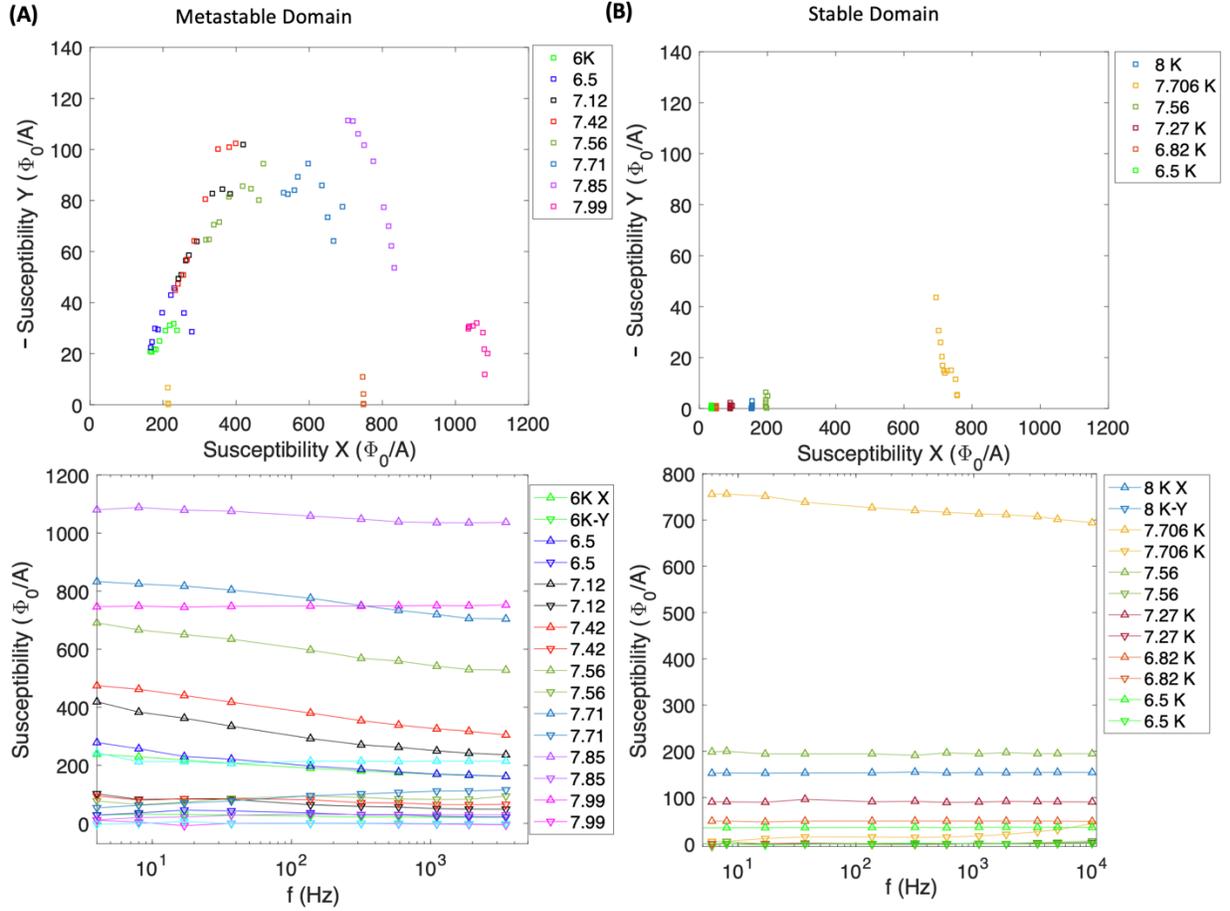

*Figure A 3. Cole-Cole plots of susceptibility above (A) metastable domains and (B) stable domains, shown in the upper panels. The lower panels show the corresponding in-phase and out-of-phase components of susceptibility as a function of frequency at fixed temperatures. The main difference between the Cole-Cole plots from stable and metastable regions is that stable regions essentially show nothing meaningful up until very close to the critical temperature, ~8 K. The main importance of the Cole-Cole plot in (A) is that within the measured frequency range, the plots at fixed temperature do not develop full arcs. Quite the opposite is the case that the arcs are very short. This observation is typical of glassy spin (or charge) systems with a very broad distribution of relaxation times [36,37,40], which, in addition to the slow fluctuations shown in Figure 5, is another strong evidence for magnetic frustration in the metastable regions.*

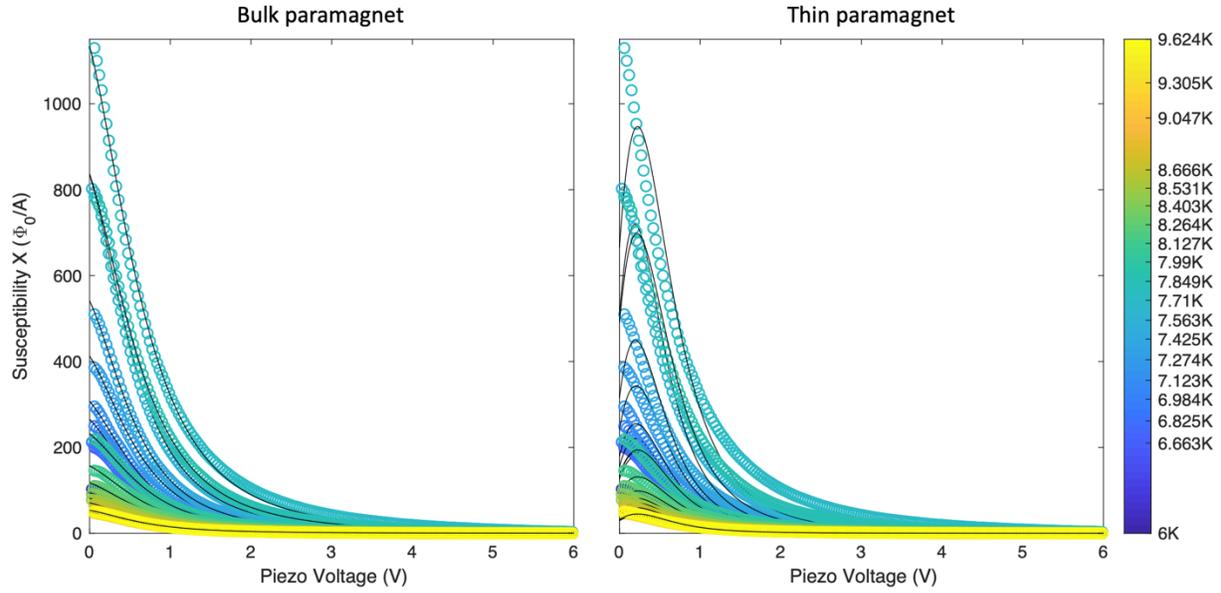

*Figure A 4. Approach curves: bulk vs. thin film fits. These are susceptibly data over a metastable region as a function of the distance of the SQUID sensor from the sample surface (proportional to the Z piezo voltage). The Kirtley-Kogan bulk paramagnet and thin film models [62] were fitted to the approach curves shown by black solid lines on the left and right panels respectively. The thin film model fails on a qualitative and quantitative level for the typical initial SQUID-sample separation of 0.8 µm. These data put a limit of the thickness of the metastable domains at $d_m \gtrsim O(1) \cdot a$, where $a = 7$ µm is the nominal radius of the field coil of the susceptometer. These data rule out the surface-only origin of the metastable domains.*

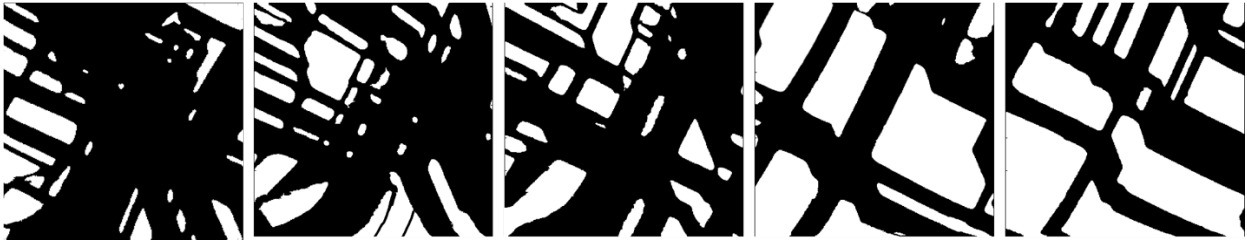

*Figure A 5. Binary masks generated at a mid-way threshold between the stable and metastable domains in the in-phase susceptibility channel. These masks are used to count the domains (e.g. Figure 2) and calculated their size using built-in Matlab ™ routines.*

Sample B

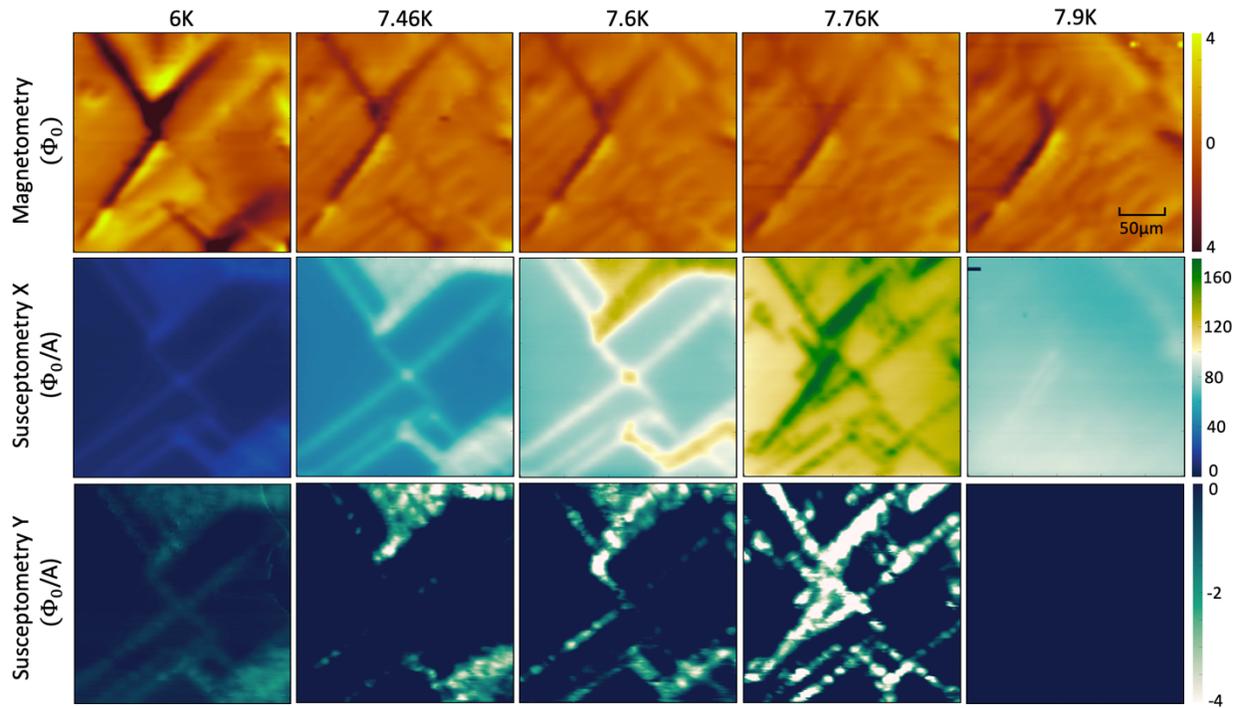

*Figure A 6. Temperature evolution of the magnetic domains in CeAlSi Sample B. Top row: dc magnetizations in flux units. Middle and bottom rows: the in-phase component and the out-of-phase component of the ac susceptibility, respectively. Similarly to Sample A, shown in the main text, dark green and dark marine areas in the in-phase and out-of-phase components, respectively, are stable in-plane domains. Light green and light marine areas in the in-phase and out-of-phase components, respectively, are metastable domains. The stable domains shrink, while the metastable grow towards the transition, until they collapse to narrow features roughly lined up with the crystallographic axes. The dc magnetization contrast reduces towards the transition, while the amplitude of the in-phase and out-of-phase signals grow. The metastable regions show enhanced signals. Above the transition temperature, all signals collapse abruptly to small values.*

## Estimates of Magnetoelastic Response

To comprehend the magnitude of the magnetoelastic response for CeAlSi, we first calculate the magnetic energy density of Ce moments, $E_M = \frac{\mathbf{B} \cdot \mathbf{M}}{2} = \frac{B^2 \cos\theta}{2\mu_0} = \frac{(0.1260 \text{ T})^2 \cos(34^0)}{2\mu_0} = 5.2 \cdot 10^3 \left(\frac{J}{m^3}\right)$, where B is the internal coarse-grained magnetic flux density evaluated based on magnetic moments per Ce atom in ref. [35]. Then we compare $E_M$ to an approximate elastic energy $E_e = \frac{Y\epsilon^2}{2} \approx \frac{120}{2} \cdot 10^9 \text{Pa} \cdot \epsilon^2 \left(\frac{J}{m^3}\right)$ (it is approximate because we don't have values of elastic constants of CeAlSi, so we use elasticity close to that of Si). By assuming $E_M = E_e$ we obtain for the expected magnetoelastic strain due to internal magnetic field $\epsilon \approx 3 \cdot 10^{-4}$.

Now, we shall compare this value to the expected value from our magnetoelasticity measurements:

$$\epsilon_{exp}(6K) = k \cdot B \approx 0.33 \cdot 10^{-6} \text{parts/Oe} \cdot 1260 \text{ Oe} \approx 4 \cdot 10^{-4}$$

Obviously, these values of $\epsilon \approx 3 \cdot 10^{-4}$ and $\epsilon_{exp} = 4 \cdot 10^{-4}$ are very close. Which means that the elastic response is not surprising in terms of the magnitude [63] of the deformation but is unique to CeAlSi in terms of the drastic local magnetic phase modification.